\documentclass[nosuperscriptaddress, showpacs, twocolumn,titlepage, showkeys, nofootinbib]{revtex4-1}
\usepackage{times}
\usepackage{amsmath}
\usepackage{amssymb}
\usepackage{graphicx}
\usepackage{subfigure}
\usepackage{booktabs}
\usepackage{rotating}
\usepackage{xcolor}
\usepackage{braket,mleftright}
\usepackage{longtable}
\usepackage{bm}
\usepackage{float}

\makeatletter

\newcommand{\Rmnum}[1]{\expandafter\@slowromancap\romannumeral #1@}
\makeatother

\begin{document}

\title{Quantum version of a generalized Monty Hall game and its possible applications to quantum secure communications}

\author{L. F. Quezada}
\email{E-mail: lfqm1987@ciencias.unam.mx}
\affiliation{Laboratorio de Informaci\'{o}n Cu\'{a}ntica, Centro de Innovaci\'on y Desarrollo Tecnol\'ogico en C\'omputo, Instituto Polit\'ecnico Nacional, UPALM, 07700, Ciudad de M\'exico, M\'exico.}

\author{Shi-Hai Dong}
\email{E-mail: dongsh2@yahoo.com}
\affiliation{Laboratorio de Informaci\'{o}n Cu\'{a}ntica, Centro de Innovaci\'on y Desarrollo Tecnol\'ogico en C\'omputo, Instituto Polit\'ecnico Nacional, UPALM, 07700, Ciudad de M\'exico, M\'exico.}

\begin{abstract}
In this work we propose a quantum version of a generalized Monty Hall game, that is, one in which the parameters of the game are left free, and not fixed on its regular values. The developed quantum scheme is then used to study the expected payoff of the player, using both a separable and an entangled initial-state. In the two cases, the classical mixed-strategy payoff is recovered under certain conditions. Lastly, we extend our quantum scheme to include multiple independent players, and use this extension to sketch two possible application of the game mechanics to quantum networks, specifically, two validated, mult-party, key-distribution, quantum protocols.
\end{abstract}

\maketitle

\section{Introduction}

Due to its versatility, the area of mathematics known as game theory, which studies the strategies to be made by rational decision-makers in a conflict situation, has been proven useful to solve problems in a wide range of areas, including economics, biology and computer science \cite{CGT1,CGT2,CGT3}. Motivated by these vast applications, and the recent development and commercial availability of quantum machines, quantum theorists start combining the game theory methodology with some features of quantum theory, such as superposition of states, projective measurements and quantum entanglement \cite{QGT1,QGT3,QGT4,QGT5,QGT6,QGT7,QGT8,QGT9,QGT10,QGT11,QGT12}, leading to the creation of quantum game theory.

Applications of quantum game theory have already been found in the study of quantum coherence \cite{QGA1} and quantum mechanics foundations \cite{QGA2}, in a quantum-like description of markets and economics \cite{QGA3} and in the development of quantum key-distribution protocols \cite{QGA4}. Furthermore, in 1999 Eisert et al. conjectured that ``survival games'' might be played by nature at a molecular level \cite{QGN1} and in 2014, Bohl et al. concluded that classical game theory can correctly model some of the behaviors of viruses, genes and proteins \cite{QGN2}, suggesting that smaller molecules can be considered as players in a quantum game. These results, along with the quantum mechanical description of protein folding \cite{QGA5}, could lead to potential applications of quantum game theory in medicine and biotechnology.

One of the games analyzed by quantum theorists was the one presented in the famous Monty Hall problem \cite{CMH1,CMH2,CMH3}, mainly because of its counter-intuitive result, which led to an intense discussion between numerous probability experts from 1975 to 1999. This game is particularly interesting from an information-theory perspective, since it illustrates how the application of a seemingly null operation over the system by one of the players, provides information on it to the other player. In 2001, using a quantum version of the Monty Hall game, Li et al. found that quantum measurements could change the expected payoffs of the players, allowing a fair two-party zero-sum game to be carried out \cite{QMHA1}. In 2016, Kurzyk and Glos constructed a quantum version of the Monty Hall problem based on a generalization of Bayesian networks, allowing them to show the difference between classical and quantum Bayesian reasoning \cite{QMHA2}.

The development of a quantum game from its classical counterpart is completely subjective, although it is usually done by replacing the strategies and procedures of the classical game with elements present in the study of a quantum system, such as states, unitary operators and projective measurements. As a result of this, there are to date various approaches and different quantum versions of the Monty Hall game \cite{QMHA1,QMH1,QMH2,QMH3,QMHA2,QMH4}.

In our previous work \cite{QGA4}, we proposed two quantum key-distribution protocols based on the quantum version of the regular Monty Hall game developed by Flitney and Abbott \cite{QMH1}. Since the regular Monty Hall game mechanics and parameters only allow the game to be played by two parties, the formalism in Flitney and Abbott's work could not be used to develop a multi-party quantum protocol. In this work we focus on developing a quantum version of a generalized Monty Hall game, that is, one in which the parameters of the game are not fixed on its regular values, and analyze the players' expected payoffs under certain strategies. From this more general quantum scheme, we recover the classical results, and analyze both a case in which the initial state of the game is separable, and one in which is entangled. Lastly, we extend our proposed scheme to include multiple independent players, and use this extension to sketch a possible application of the game mechanics to quantum networks.

The paper is organized as follows: A summary of the classical Monty Hall game results, as well as their generalization, are given in Sec. \ref{cMH}. In Sec. \ref{qMH} we introduce the formalism of our proposed quantum scheme of the generalized game, and present the results in section \ref{res}. The separable case is studied in Subsec. \ref{nent}, while the entangled case is studied in Subsec. \ref{ent}. Lastly, in Subsec. \ref{multi} we expose a multi-player version of our proposed quantum scheme, while in Subsecs. \ref{direct} and \ref{motivated} we sketch two possible applications of the multi-player game to quantum networks.

\section{Classical Monty Hall game} \label{cMH}

The Monty Hall problem, where the Monty Hall game is presented, is a puzzle in probability theory proposed in 1975 \cite{CMH1,CMH2} and actively discussed in 1991 \cite{CMH3} due to its counter-intuitive result.

The mechanics of the game is as follows: In a contest, a player is asked to choose between three doors, behind one of which is a car (the prize) and the other two are goats. If the door chosen by the player is the one with the car behind, the host of the contest (Monty Hall), who knows where the car is, randomly opens one of the other two doors, revealing to the player one of the goats. If the door chosen by the player is the one with a goat behind, the host opens the only remaining door hiding a goat. The player is then asked if he wishes to open his initially chosen door or prefers to switch and open the other remaining closed door. It is found that the best strategy for the player to follow is to switch, as the probability of finding the car behind the door he initially chose is $1/3$, while the probability of finding the car behind the other door is $2/3$.

A simple but rigorous way to obtain the above result is the following: Both the location of the car and the initially chosen door by the player are random, independent events with a probability of $1/3$ each. This means that the probability of simultaneously the car being behind door $j$ and the player initially choosing door $i$ is simply $P(i,j) = 1/9$ for all $i,j = 1,2,3$. Therefore, all the nine events of the corresponding sample space $\left\lbrace (i,j) \, | i,j = 1,2,3  \right\rbrace$, have a probability of $1/9$. Notice that the events of the form $(i,j)$ with $i = j$, are the cases in which the player wins if he decides to open his initially chosen door, while the events of the form $(i,j)$ with $i\neq j$, represent the cases in which the player wins if he decides to switch. The probability $P_{ns}$ of the player finding the prize by opening his initially chosen door, is calculated by adding the probabilities of the elements corresponding to that event:
\begin{align}
P_{ns} = P(1,1) + P(2,2) + P(3,3) = \frac{1}{3}. \label{eq:a}
\end{align}
Analogously, the probability $P_{s}$ of the player finding the prize by switching doors is
\begin{align}
P_{s} = \, &P(1,2) + P (1,3) + P(2,1) \notag \\ & + P(2,3) + P(3,1) + P(3,2) = \frac{2}{3}.\label{eq:b}
\end{align}

A possible generalization of the Monty Hall game can be made by considering an arbitrary number of doors $d$ to hide the prize and various not-prized doors to be opened by the host. In this case, the sample space corresponding to simultaneously the car being behind door $j$ and the player initially choosing door $i$ has the form $\left\lbrace (i,j) \, | i,j = 1,\dots,d  \right\rbrace$, and each element has a probability of $1/d^{2}$. Notice again that the $d$ events of the form $(i,j)$ with $i = j$, are the cases in which the player wins if he decides to open his initially chosen door, this means that the probability $P_{ns}$ of the player finding the prize by opening his initially chosen door is

\begin{equation}
P_{ns} = \frac{1}{d}. \label{eq:ga}
\end{equation}
On the other hand, the events of the form $(i,j)$ with $i \neq j$ have the same probability of $1/d^{2}$ to happen, but not of the player to win by switching. That can be calculated by considering that, after the host opens $m$ not-prized doors, there are $d-m-1$ possible doors for the player to switch, meaning that each of the $d(d-1)$ events of the form $(i,j)$ with $i \neq j$, has a probability of $1/d^{2}(d-m-1)$ of the player winning by switching. Thus, the probability of the player finding the prize by switching doors is

\begin{equation}
P_{s} = \left( \frac{d-1}{d-m-1}\right)  \frac{1}{d}. \label{eq:gb}
\end{equation}

The game can be further generalized to be a multi-player one. Consider $n$ players (counting the host as a player) and suppose that all of them win the prize if they open the correct door, in this way, the players' strategies are independent from each other's, and expressions \eqref{eq:ga} and \eqref{eq:gb} remain the same for each one of them.

For consistency with the game mechanics, in this generalization scheme, the parameters $d$ (the total number of doors), $m$ (the not-prized doors to be opened by the host) and $n$ (the number of players including the host) are subject to the following restrictions:

\begin{align}
\label{c1} n \geq 2, \\
\label{c2} d-n \geq m \geq 0.
\end{align}

It is worth mentioning that $P_{s} + P_{ns} = 1$ only when $d-n = m$.

\section{Quantum version of the generalized Monty Hall game} \label{qMH}

In this section we propose a quantum version of the generalized Monty Hall game presented in section \ref{cMH}. As it is usual in both quantum information and quantum game theory, we use the characters Alice and Bob as the host and the player respectively.

The states of the game will be described by normalized vectors $\Ket{\psi}$ living in the space
\begin{equation}
\label{space} \mathcal{H} = \bigotimes^{m+n}_{i=1} \mathcal{H}_{i} ,
\end{equation}
where $m$ is the number of not-prized doors to be opened by Alice (the host), $n$ is the number of players (counting the host as a player) and every $\mathcal{H}_{i}$ is the complex vector space of dimension $d$ (the total number of doors). Without loss of generality, and in order to simplify the game description, here we focus on the case $n=2$ (i.e. the host and one player). Thus, we write a state of the game in the space $\mathcal{H}$ as
\begin{equation}
\Ket{\psi} = \Ket{\vec{o}, \, b, \, a} = \Ket{o_{m}, \dots, o_{1}, \, b, \, a},
\end{equation}
where $a \in \left\lbrace 0,\dots, d-1 \right\rbrace $ indicates the door in which Alice hides the prize, $b \in \left\lbrace 0,\dots, d-1 \right\rbrace $ the door chosen by Bob and $o_{i} \in \left\lbrace 0,\dots, d-1 \right\rbrace$ the not-prized doors to be opened by Alice.

The game mechanics creates some restrictions on the labels $a,b,o_{i}$. As we mentioned in section \ref{cMH}, the location of the prize and the initial chosen door by the player are independent events, this means $a$ and $b$ can have any value in $\left\lbrace 0,\dots , d-1 \right\rbrace $. However, in order for Alice to open $m$ different not-prized doors, the labels $o_{i}$ must be different from each other, that is, $o_{i} \neq o_{j}$ if $i \neq j$. Furthermore, the opened not-prized doors must also be different from the one where the prize is hidden and from the one selected by Bob, that is, $o_{i} \neq a,b$ for all $i$. We denote by $\mathcal{G}$ the subspace of $\mathcal{H}$ generated by all possible game states.

We consider the initial state of the game to be of the form
\begin{equation}
\label{initstate} \Ket{\psi^{(i)}} = \Ket{\vec{0}} \otimes \Ket{\phi^{(i)}},
\end{equation}
where $\Ket{\phi^{(i)}}$ is any state in the space corresponding to the prize's location and Bob's initially chosen door (labels $a$ and $b$), and $\vec{0}$ accounts for all the opened doors labels $o_{i}$ to be initially $0$.

The game begins with Alice and Bob applying their strategies on the initial state $\Ket{\phi^{(i)}}$. That is, Alice hides the prize behind a door by applying a special unitary operator $\hat{A}$ on the first qudit (the one labeled as ``$a$''), and Bob chooses a door by applying a special unitary operator $\hat{B}$ on the second qudit (the one labeled as ``$b$''). The next step is for Alice to open $m$ different not-prized doors, which is implemented by applying in succession the door-opening operators:
\begin{multline}
\label{open} \hat{O}_{j} = \\ \sum^{}_{\vec{o}_{j},b,a} \frac{\varepsilon\left( a,b,\vec{o}_{j} \right)}{\sqrt{d+1-j-\text{U}(a,b)}} \cdot \Ket{o_{j},\vec{o}_{j-1},b,a} \Bra{0,\vec{o}_{j-1},b,a},
\end{multline}
where the sum runs from $0$ to $d-1$, $j \in \left\lbrace 1,\dots,m \right\rbrace$, $\vec{o}_{k}$ stands for the ordered set of labels $(o_{k},o_{k-1},\dots,o_{1})$,
\begin{equation}
\varepsilon\left( \vec{x} \right) = \left\lbrace
\begin{tabular}{ccc}
	$0$ & $\quad$ &  if any two labels in $\vec{x}$ \\
	 &  & have the same value,  \\
	 &  & \\
	 $1$ & $\quad$ & otherwise, \\
\end{tabular}
\right.
\end{equation}
and the $\text{U}$ function returns the number of unique elements in its argument. For example, $\text{U}(1,3,5) = 3$, while $\text{U}(0,2,2) = 2$.

Each of the possible $m$ door-opening operators $\hat{O}_{j}$, acts on the space corresponding to the labels $a,b,o_{1},\dots,o_{j}$, creating a superposition of all the possible doors to be opened, namely, the ones that remain after Alice and Bob have played their strategies and the ones that have not already been opened by the previous operators $\hat{O}_{k}$ with $k<j$. Furthermore, every $\hat{O}_{j}$ is a special unitary operator in its domain (states with $\vec{o}_{j} = 0$) and thus can be arbitrarily extended to be special unitary in all $\mathcal{H}$.

Lastly, Bob decides whether he keeps his initial choice or prefers to switch it. The switching case can be implemented by applying the door-switching operator
\begin{equation}
\label{switch} \hat{S} = \sum^{}_{\vec{o},b} \varepsilon\left( b,\vec{o} \right)  \cdot \Ket{\vec{o},b\oplus \ell_{b,\vec{o}} } \Bra{\vec{o},b},
\end{equation}
where the sum runs again from $0$ to $d-1$, $b\oplus \ell_{b,\vec{o}}$ symbolizes the sum mod. $d$ and
\begin{equation}
\label{ell} \ell_{b,\vec{o}} = \underset{k \in \left\lbrace 1,\dots,d-1 \right\rbrace }{\min} \left\lbrace k \, | \, (b \oplus k) \notin \vec{o} \right\rbrace.
\end{equation}

The door-switching operator $\hat{S}$ acts on the space corresponding to the labels $b,o_{1},\dots,o_{m}$, changing label $b$ (Bob's initially chosen door) to the next one (mod. $d$) available, that is, a door different from all the doors that have already been opened by the operators $\hat{O}_{j}$. Notice that $\hat{S}$ is also special unitary in its domain and hence it can be extended to be in all $\mathcal{H}$.

We introduce a parameter $\gamma \in \left[ 0,\pi/2 \right]$ to account for a quantum mixed strategy approach in the switching decision. This is implemented by Bob applying the operator
\begin{equation}
\label{qmix} \cos\gamma \, \hat{I}_{d^{m+1}} + \sin\gamma \, \hat{S},
\end{equation}
where $\hat{I}_{d^{m+1}}$ is the identity operator of dimension $d^{m+1}$. The factor $\sin\gamma$ represents the probability amplitude of Bob applying the door-switching operator $\hat{S}$, while $\cos\gamma$ represents the probability amplitude of Bob keeping his initial choice.

The final state of the game is therefore given by
\begin{align}
\label{psif} \Ket{\psi^{(f)}} &= \notag \\
&\left[ \left( \cos\gamma \, \hat{I}_{d^{m+1}} + \sin\gamma \, \hat{S}\right) \otimes \hat{I}_{d} \right]  \cdot \notag \\ & \left[ \prod^{1}_{j=m} \left( \hat{I}_{d^{m-j}} \otimes \hat{O}_{j} \right) \right]  \cdot \left( \hat{I}_{d^{m}} \otimes \hat{B} \otimes \hat{A} \right) \, \Ket{\psi^{(i)}}.
\end{align}
Bob wins if he opens the same door that Alice chose to hide the prize, that is, Bob wins if his qudit has the same value as Alice's. Thus the probability of Bob winning the game, i.e. his expected payoff, is
\begin{equation}
\label{payoff} \left\langle \$_{B} \right\rangle =  \sum^{}_{i,\vec{o}} \left| \Braket{\vec{o},i,i | \psi^{(f)}} \right|^{2},
\end{equation}
where, once again, the sum runs from $0$ to $d-1$.

Furthermore, as a non-cooperative game, Alice wins if Bob fails to choose the correct door. Therefore, her expected payoff is simply $\left\langle \$_{A} \right\rangle = 1 - \left\langle \$_{B} \right\rangle$.

It is worth mentioning that the quantum mixed strategy modeled by the parameter $\gamma$ in equation \eqref{qmix} creates a superposition of the switching and not-switching cases, and thus is essentially different from a classical mixed strategy, where the expected payoff is given by $\$ = \cos^{2}\gamma \, \$_{not-switch} + \sin^{2}\gamma \, \$_{switch}$.

\section{Results} \label{res}
\subsection{Without entanglement} \label{nent}

\begin{figure}[!t]
\includegraphics[width=0.9\linewidth]{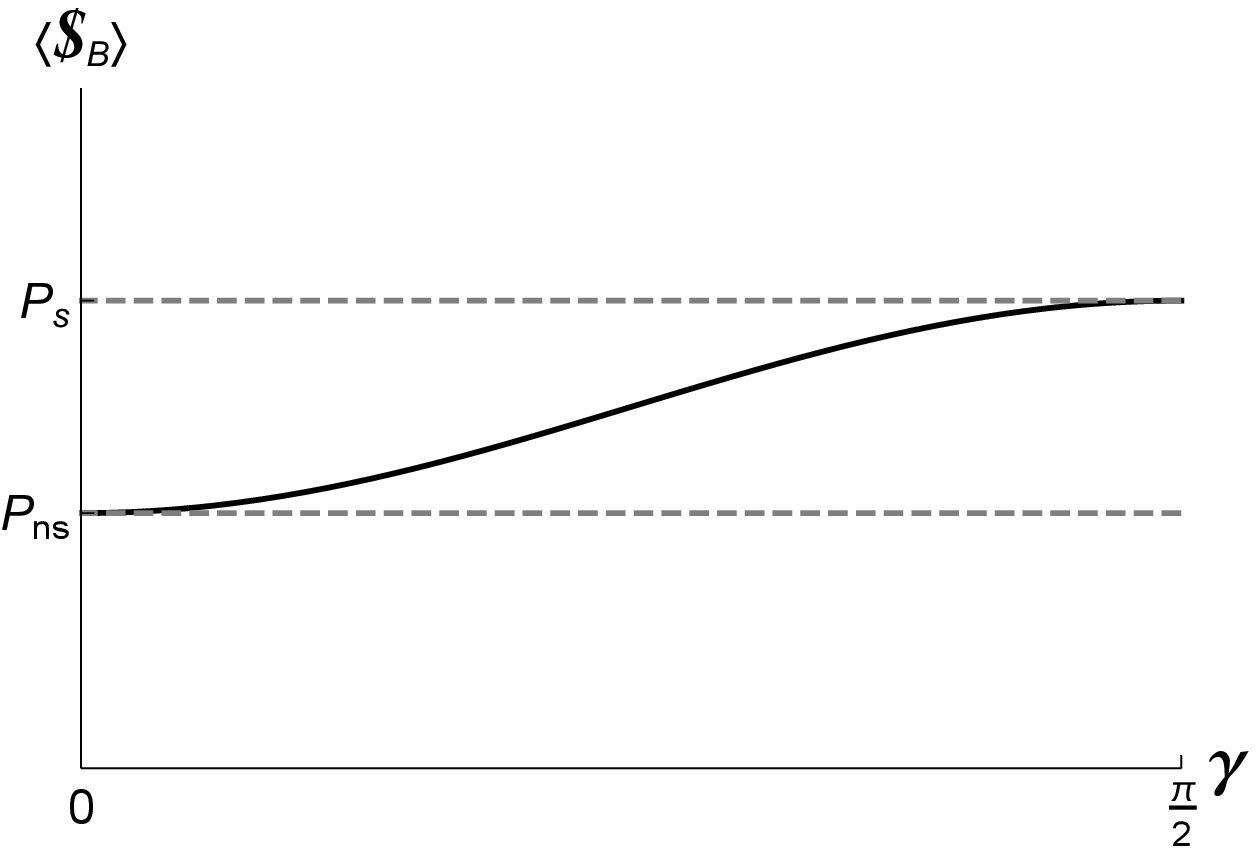}
\caption{\label{Nentc} Classical-mixed-strategy's behavior of Bob's expected payoff $\left\langle \$_{B} \right\rangle$ as a function of the parameter $\gamma$, obtained with a non-entangled initial state.}
\end{figure}

In this section we analyze the expected payoff of the player (Bob) when the initial state of the game \eqref{initstate} is not entangled. Specifically, as usual in quantum game theory and quantum computation, we consider the initial state to be
\begin{equation}
\Ket{\psi^{(i)}} = \Ket{\vec{0}} \otimes \Ket{00},
\end{equation}
where $a$, $b$ and all labels in $\vec{o}$ are initialized at zero.

In this case, using equations \eqref{psif} and \eqref{payoff}, and identifying the player's strategies $\hat{A}$, $\hat{B}$ with their respective matrix elements $a_{i,j}$, $b_{i,j}$, the expected payoff of Bob is found to be
\begin{multline}
\label{nepay} \left\langle \$_{B} \right\rangle = \frac{(d-m-1)!}{(d-1)!} \cdot \sum^{}_{j,\vec{o}} \left| a_{j,0} \right|^{2} \cdot \left| \vphantom{\sqrt{\frac{d-1}{d-m-1}}} \cos\gamma \cdot b_{j,0} \cdot \varepsilon\left(\vec{o},j \right) \right. \\ \left. + \sqrt{\frac{d-1}{d-m-1}} \, \sin\gamma \cdot b_{j \ominus \lambda_{j,\vec{o}},0} \cdot \varepsilon\left( \vec{o}, j \ominus \lambda_{j,\vec{o}}, j \right)    \right|^{2},
\end{multline}
where the sum runs from $0$ to $d-1$, the symbol $\ominus$ symbolizes the subtraction mod. $d$ and the term $\lambda_{j,\vec{o}}$ is defined as
\begin{equation}
\lambda_{j,\vec{o}} = \underset{k \in \left\lbrace 1,\dots,d-1 \right\rbrace }{\min} \left\lbrace k \, | \, (j \ominus k) \notin \vec{o} \right\rbrace.
\end{equation}

Let us analyze some specific cases. The classical mixed strategy can be recovered from the proposed quantum scheme by considering the strategies $\hat{A} = \text{QFT}$ and $\hat{B} = \text{SUM}_{d}(i)$, where QFT stands for the quantum Fourier transform and $\text{SUM}_{d}(i)$ is the sum of $i$ mod. $d$, that is
\begin{equation}
\label{sumdi} \text{SUM}_{d}(i) = \sum^{d-1}_{j=0} \, \Ket{j \oplus i}\Bra{j}.
\end{equation}
These strategies lead to the matrix elements in expression \eqref{nepay} to respectively be
\begin{align}
a_{k,0} &= \frac{1}{\sqrt{d}}, \\
b_{k,0} &= \delta_{ik},
\end{align}
and represent the case in which Alice hides the prize in a homogeneous superposition of doors and Bob chooses a specific one. Figure \ref{Nentc} shows the classical-mixed-strategy's behavior of the expected payoff $\left\langle \$_{B} \right\rangle$ as a function of the parameter $\gamma$, which is given by
\begin{equation}
\label{nepayc} \left\langle \$_{B} \right\rangle = P_{ns} \cos^{2}\gamma + P_{s} \sin^{2}\gamma.
\end{equation}
Notice that the mixed classical payoff in Figure \ref{Nentc} does not exceed the classical probability of winning by switching $P_{s}$ for any value of $\gamma$.

\begin{figure}[!t]
\centering
\includegraphics[width=0.9\linewidth]{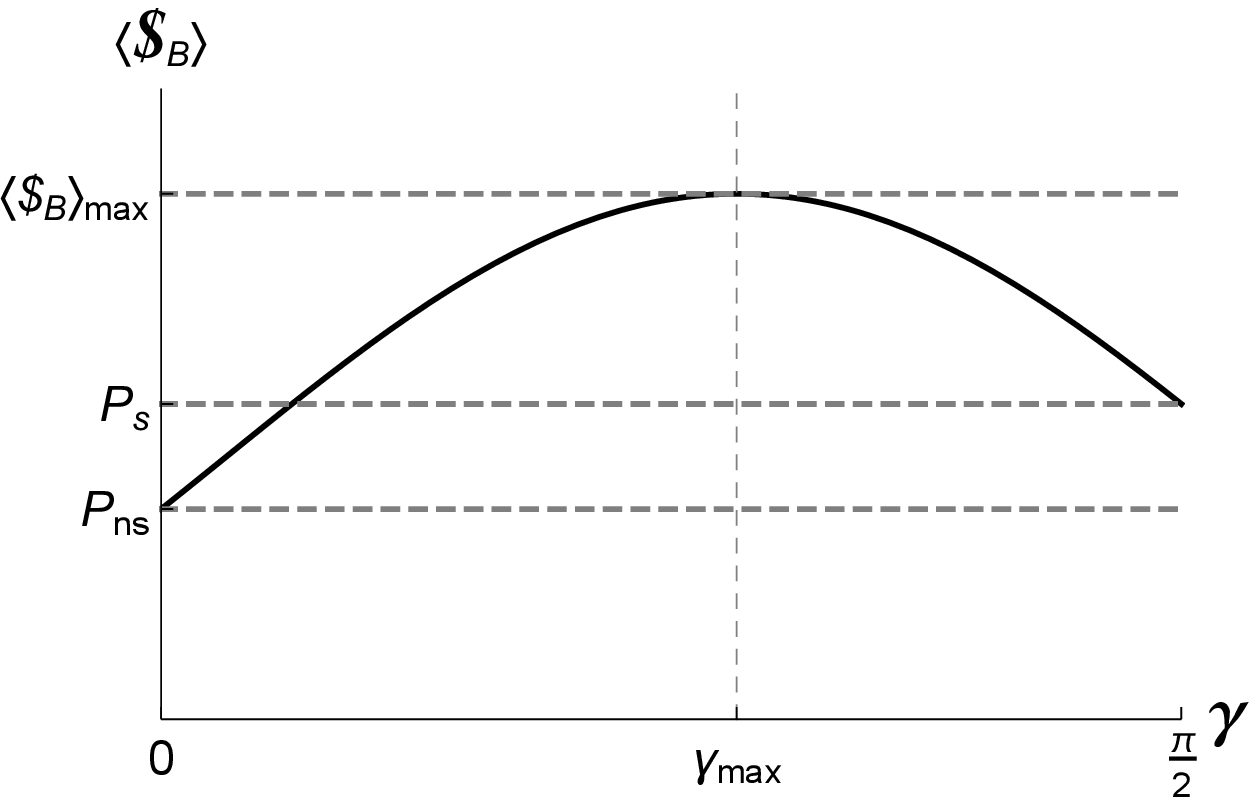}
\caption{\label{Nent} Bob's expected payoff $\left\langle \$_{B} \right\rangle$ as a function of the parameter $\gamma$, obtained with a non-entangled initial state and using $\hat{B} = \text{QFT}$. The curve is independent of Alice's strategy.}
\end{figure}

A particularly interesting case is where Bob's strategy is to choose a homogeneous superposition of all possible doors, that is, choosing $\hat{B} = \text{QFT}$, which leads to $b_{k,0} = 1/\sqrt{d}$ for all $k$. This results in Bob's expected payoff in equation \eqref{nepay} not depending on Alice's strategy, and having the form
\begin{equation}
\label{nepayq} \left\langle \$_{B} \right\rangle = \left| \sqrt{P_{ns}} \cos\gamma + \sqrt{P_{s}} \sin\gamma \right|^{2}.
\end{equation}
Figure \ref{Nent} shows the plot of expression \eqref{nepayq} as a function of the parameter $\gamma$. It can be seen that, in this case, $\left\langle \$_{B} \right\rangle$ has a maximum value of
\begin{equation}
\left\langle \$_{B} \right\rangle_{max} = \left| \frac{P_{ns}}{\sqrt{P_{ns}+P_{s}}} + \frac{P_{s}}{\sqrt{P_{ns}+P_{s}}} \right|^{2}
\end{equation}
at
\begin{equation}
\gamma_{max} = \arctan{\sqrt{\frac{P_{s}}{P_{ns}}}},
\end{equation}
which is greater than $P_{s}$ (the maximum value attainable by a classical strategy), meaning that, in this particular game, access to quantum strategies leads to a greater payoff. Furthermore, if $P_{ns} + P_{s} = 1$, then $\left\langle \$_{B} \right\rangle_{max} = 1$. This means that, by choosing the strategies $\hat{B} = \text{QFT}$ and $\gamma = \gamma_{max}$, Bob can win every time regardless of Alice's strategy. Therefore, in this latter case, the set of strategies $\left\lbrace \hat{A} \in \text{SU(\textit{d})}, \hat{B} = \text{QFT}, \gamma = \gamma_{max} \right\rbrace $ conforms a quantum weak Nash equilibrium.

\begin{figure}[!t]
\centering
\includegraphics[width=0.9\linewidth]{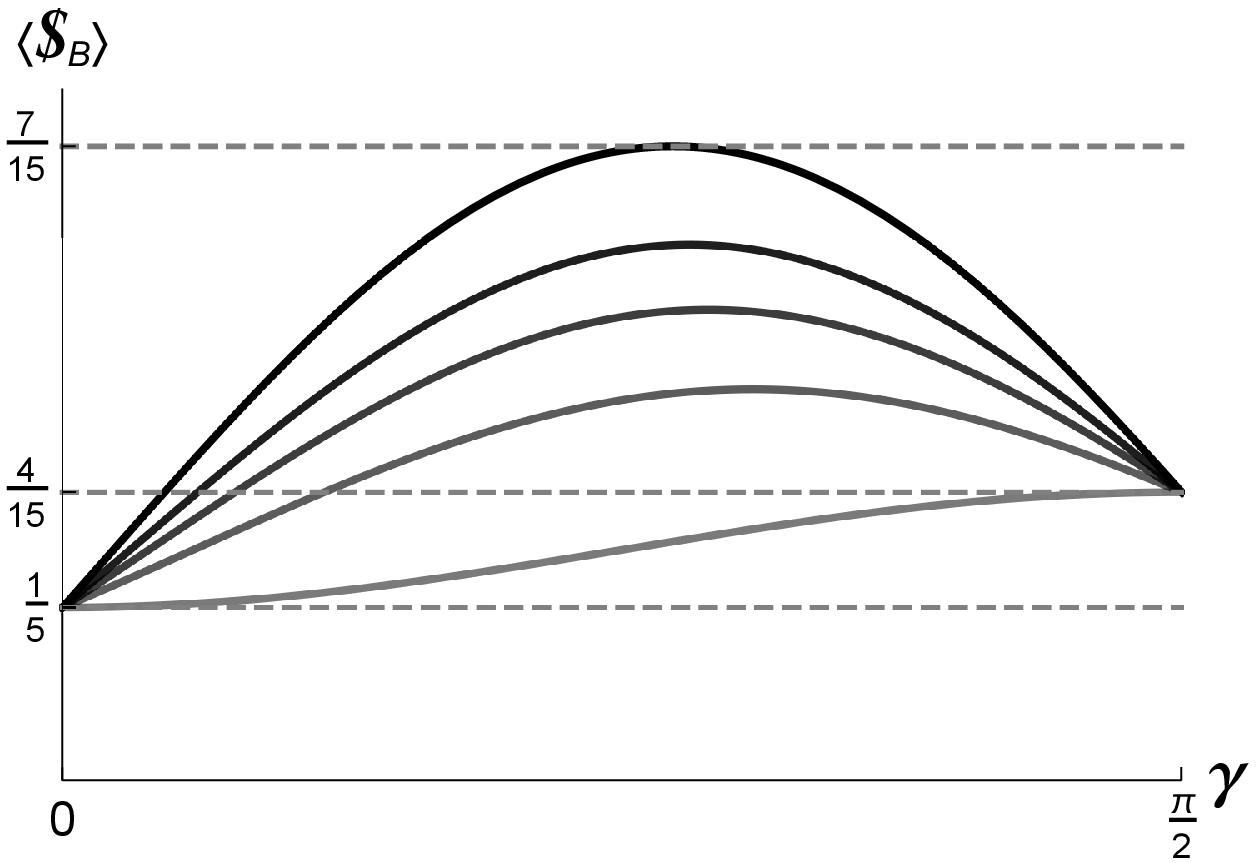}
\caption{Bob's expected payoff $\left\langle \$_{B} \right\rangle$ as a function of the parameter $\gamma$, obtained using $d=5$ and $m=1$, with a non-entangled initial state and applying (bottom-up) a classical mixed strategy, a homogeneous superposition of two states, three states, four states and $\hat{B} = \text{QFT}$. \label{Nentfam}}
\end{figure}

The graphics in Figures \ref{Nentc} and \ref{Nent} can be seen as the extreme cases of the family of curves obtained by varying Bob's strategy when Alice's strategy is fixed at $\hat{A} = \text{QFT}$. Figure \ref{Nentfam} shows, for the case $d = 5$, $m = 1$, a subset of this family of curves when $\hat{A} = \text{QFT}$, and (bottom up) $\hat{B} \ket{0} = \ket{0}$, $\hat{B} \ket{0} = \frac{1}{\sqrt{2}} \left(  \ket{0} + \ket{1} \right) $, $\hat{B} \ket{0} = \frac{1}{\sqrt{3}} \left(  \ket{0} + \ket{1} + \ket{2} \right)$, $\hat{B} \ket{0} = \frac{1}{\sqrt{4}} \left(  \ket{0} + \ket{1} + \ket{2} + \ket{3} \right)$ and $\hat{B} = \text{QFT}$. We can see from expression \eqref{nepay} that, the more doors are included in the superposition of Bob's strategy (more $b_{j,0} \neq 0$), the closer the curve is going to be to the case where $\left\langle \$_{B} \right\rangle_{max}$ is reached.

\subsection{With entanglement between host and player} \label{ent}

In this section we analyze the expected payoff of the player (Bob) when the initial state between him and the host (Alice), $\Ket{\phi^{i}}$ in equation \eqref{initstate}, is entangled. Specifically, we consider the state $\Ket{\phi^{i}}$ to be the GHZ state \cite{GHZ} of dimension $d$ between two parties:
\begin{equation}
\Ket{\phi^{(i)}} = \sum_{j=0}^{d-1} \, \Ket{jj},
\end{equation}
leading the initial state of the game \eqref{initstate} to be
\begin{equation}
\label{psiient} \Ket{\psi^{(i)}} = \Ket{\vec{0}} \otimes \sum_{j=0}^{d-1} \, \Ket{jj}.
\end{equation}

In this case, using equations \eqref{psif} and \eqref{payoff}, and identifying the player's strategies $\hat{A}$, $\hat{B}$ with their respective matrix elements $a_{i,j}$, $b_{i,j}$, the expected payoff of Bob is found to be
\begin{multline}
\label{epay} \left\langle \$_{B} \right\rangle = \frac{(d-m-1)!}{d!} \cdot \sum^{}_{j,\vec{o}} \left| \vphantom{\sqrt{\frac{d-1}{d-m-1}}} \cos\gamma \cdot \varepsilon\left(\vec{o},j \right) \cdot \sum_{i} \left( a_{j,i} \cdot  b_{j,i} \right)  \right. \\ \left. + \sqrt{\frac{d-1}{d-m-1}} \, \sin\gamma \cdot \varepsilon\left( \vec{o}, j \ominus \lambda_{j,\vec{o}}, j \right) \cdot \right. \\ \left. \sum_{i} \left(  b_{j \ominus \lambda_{j,\vec{o}},i} \cdot a_{j,i} \right)  \right|^{2},
\end{multline}
where the sums run from $0$ to $d-1$.

\begin{figure}[!t]
\centering
\includegraphics[width=0.9\linewidth]{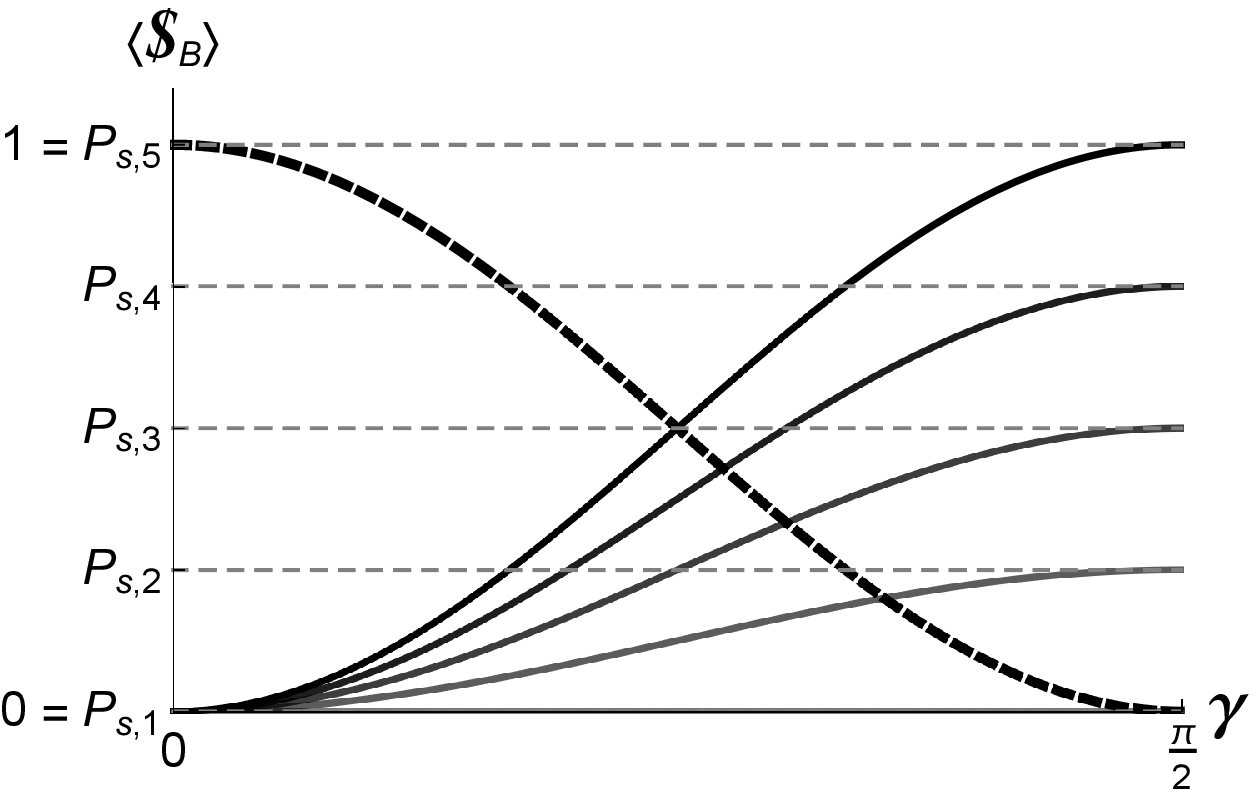}
\caption{Bob's expected payoff $\left\langle \$_{B} \right\rangle$ as a function of the parameter $\gamma$, obtained using $d=6$ and $m=3$, with an entangled initial state and applying a displacement between Alice and Bob' labels of $k=0$ (dashed line) and (bottom up) $k=1,2,3,4,5$ (continuous lines). \label{Entfam}}
\end{figure}

A first interesting result is obtained when both Alice and Bob apply the quantum Fourier transform, namely
\begin{equation}
a_{j,k} = b_{j,k} = \frac{1}{\sqrt{d}} e^{\frac{2 \pi i }{d} j k}.
\end{equation}
In contrast with the obtained expected payoff in the non-entangled case \eqref{nepayq}, the extra phase in the matrix elements leads to the terms that combine the cosine and the sine in expression \eqref{epay}, to cancel, resulting in the entangled expected payoff to be the same as the classical one, that is, the one in equation \eqref{nepayc}.

The GHZ state perfectly correlates Alice and Bob's labels, meaning that if neither Alice nor Bob applies a strategy ($\hat{A} = \hat{B} = \hat{I}_{d}$), Bob will win with a probability of $1$ if he decides not to apply the door-switching operator, and with a probability of $0$ if he decides to apply it. The above can be regarded as a particular case of both labels being displaced by a fixed amount $k \in \left\lbrace 0,\dots,d-1 \right\rbrace $, which can be implemented by Alice and Bob respectively applying the strategies introduced in equation \eqref{sumdi}, specifically $\hat{A} = \text{SUM}_{d}(i)$ and $\hat{B} = \text{SUM}_{d}(i+k)$ for any $i$. In this case, the expected payoff is found to be
\begin{equation}
\label{dispay} \left\langle \$_{B} \right\rangle = P_{ns,k} \cos^{2}\gamma + P_{s,k} \sin^{2}\gamma,
\end{equation}
where
\begin{equation}
P_{ns,k} =
\begin{cases}
1 \quad \text{if } k = 0,\\
0 \quad \text{if } k \neq 0,
\end{cases}
\end{equation}
\begin{equation}
P_{s,k} = \left\lbrace
\begin{tabular}{ccc}
$\displaystyle\frac{m! \, (k-1)!}{(m+k+1-d)! \, (d-2)!}$ & $\quad$ &  if $k \geq d-m-1$, \\
&  & \\
$0$ & $\quad$ & if $k < d-m-1$. \\
\end{tabular}
\right.
\end{equation}

Figure \ref{Entfam} shows, for the case $d=6$ and $m=3$, Bob's expected payoff obtained using a displacement of $k=0,1,2,3,4,5$. Notice that $\left\langle \$_{B} \right\rangle$ in equation \eqref{dispay} strongly depends on the specific displacement $k$, as well as in the difference between the total number of doors $d$ and the number of doors to be opened $m$. This is due to how the door-opening operators $\hat{O}_{j}$ and the door-switching operator $\hat{S}$ work. The door-opening operators will begin to ``fill the gaps'' between Alice and Bob's displaced labels, allowing the door-switching operator to ``jump'' from Bob's initial label to Alice's one, resulting in a greater probability of winning by switching when the displacement is closest (from below) to $d$, being the extreme cases $k=0$ (perfect correlation) and $k=d-1$, where there are no gaps to fill and the door-switching operator automatically switches Bob's label to Alice's one. However, if the number of doors to be opened $m$ is not large enough to fill the gaps created by the displacement $k$, i.e. $m < d-k-1$, the door-switching operator will not be able to jump from Bob's label to Alice's, resulting in Bob's expected payoff to be zero regardless of the value of $\gamma$.

In general, as Flitney and Abbott mentioned in their quantum scheme of the regular game \cite{QMH1}, if the initial state is the one in equation \eqref{psiient}, then the game does not have a Nash equilibrium among pure quantum strategies. This is due to a useful property of the GHZ state, namely \cite{QGT3,QMH1}
\begin{equation}
\label{GHZproperty} \left( \hat{U^{*}} \otimes \hat{U} \right) \sum_{j=0}^{d-1} \, \Ket{jj} = \sum_{j=0}^{d-1} \, \Ket{jj},
\end{equation}
where $\hat{U} \in$ SU($d$) and $\hat{U^{*}}$ is its complex conjugate. In this context, expression \eqref{GHZproperty} can be regarded as the existence of a counter-strategy $\hat{U^{*}}$ for every strategy $\hat{U}$.

\section{Multi-player game and two possible applications to quantum networks} \label{Mgame}

In the last two sections we have discussed the details of a quantum scheme for the generalized two-parties (host and one player) Monty Hall game. Here we extend the formalism presented in Sec. \ref{qMH} to include multiple independent players and propose two possible applications to the area of quantum secure communications.

\subsection{Multi-player game} \label{multi}

In the multi-player case, a state of the game is written as
\begin{equation}
\Ket{\psi} = \Ket{\vec{o}, \, \vec{p} \,} = \Ket{o_{m}, \dots, o_{1}, p_{n}, \dots, p_{1}},
\end{equation}
where $p_{1}$ indicates the door in which the host (which we will refer to as $\mathcal{P}_{1}$) hides the prize, $p_{2}, \dots, p_{n}$ respectively represent the chosen doors by each of the $n-1$ players (which we will refer to as $\mathcal{P}_{2},\dots,\mathcal{P}_{n}$) and $o_{i}$ are the not-prized doors to be opened by $\mathcal{P}_{1}$.

As in the $n=2$ case, every party in $\left\lbrace \mathcal{P}_{1},\dots,\mathcal{P}_{n}\right\rbrace$ plays the game by applying a strategy $\hat{P}_{1},\dots,\hat{P}_{n} \in \text{SU}(d)$ to its own qudit. The generalization of the door-opening operators defined in equation \eqref{open} is straightforward, and is done by extending the domain of the operators to take into account all the players' labels in $\vec{p}$:
\begin{multline}
	\hat{O}_{j} = \\ \sum^{}_{\vec{o}_{j},\vec{p}} \frac{\varepsilon\left( \vec{p},\vec{o}_{j} \right)}{\sqrt{d+1-j-\text{U}(\vec{p})}} \cdot \Ket{o_{j},\vec{o}_{j-1},\vec{p} \,} \Bra{0,\vec{o}_{j-1},\vec{p} \,}.
\end{multline}
Furthermore, since every player is independent from each other, they all need a door-switching operator of their own, namely
\begin{equation}
\label{switchm} \hat{S}_{k} = \sum^{}_{\vec{o},p_{k}} \varepsilon\left(\vec{o}, p_{k} \right)  \cdot \Ket{\vec{o}, p_{k} \oplus \ell_{p_{k},\vec{o}} } \Bra{\vec{o},p_{k}},
\end{equation}
with $k\in \left\lbrace 2,\dots , n \right\rbrace $. Each operator $\hat{S}_{k}$ acts on the space corresponding to the labels $o_{1},\dots,o_{m}, p_{k}$, i.e. it only switches the door initially chosen by player $\mathcal{P}_{k}$ and does not consider the doors chosen by other players at all. Hence, the results from the previous sections apply the same to each player.

\subsection{Direct application} \label{direct}

Using the new definitions of the previous subsection, we are now in a position to sketch a direct application of the game, using the same operators and mechanics. We first propose a validated multi-party quantum key-distribution protocol in a trusted network, in which the host ($\mathcal{P}_{1}$) acts as a central node and distributes the key to the $n-1$ players $\mathcal{P}_{2},\dots,\mathcal{P}_{n}$, while other $m$ parties, which we will denote as $\mathcal{V}_{1},\dots , \mathcal{V}_{m}$, validate the distribution. It is worth mentioning that the protocol proposed in this subsection is not intended to be a secure way of distributing a random key, but rather a secure way of validating the distribution.

The protocol is developed under the condition $d = m + 2$ and goes as follows:

\begin{enumerate}
	\setcounter{enumi}{0}
	\item $\mathcal{P}_{1}$ generates the state
	\begin{equation}
	\Ket{\psi} = \Ket{\vec{0},\vec{0}} ,
	\end{equation}
	i.e. the state where all the $m$ labels in $\vec{o}$ and all the $n$ labels in $\vec{p}$ are equal to zero.
	
	\item $\mathcal{P}_{1}$ sends the qudits $p_{2},\dots,p_{n}$ to the respective $n-1$ players $\mathcal{P}_{2},\dots,\mathcal{P}_{n}$, and keeps the qudits $p_{1}$ and all the ones in $\vec{o}$.
	
	\item Each participant $\mathcal{P}_{k}$ ($k=1,\dots,n$) randomly applies, to its own qudit, one of the following two strategies: $\hat{P}_{k} = \text{SUM}_{d}(0)$, $\hat{P}_{k} = \text{SUM}_{d}(1)$. And privately stores in a classical bit $b_{k}$ its selection: $b_{k} = 0$ for $\text{SUM}_{d}(0)$ and $b_{k} = 1$ for $\text{SUM}_{d}(1)$.
	
	\item All the players send back the qudits $p_{2},\dots,p_{n}$ to $\mathcal{P}_{1}$.
	
	\item $\mathcal{P}_{1}$ sends all the state to the first validation party $\mathcal{V}_{1}$.
	
	\item $\mathcal{V}_{1}$, based on some authentication or relevant information related to the transaction, applies (or not) its own door-opening operator, and sends the state to the next validation party $\mathcal{V}_{2}$. This procedure continues until all validation parties $\mathcal{V}_{1},\dots , \mathcal{V}_{m}$ have decided to validate (apply the corresponding door-opening operator) or not the key distribution.
	
	\item $\mathcal{V}_{m}$ sends back the full state to $\mathcal{P}_{1}$.
	
	\item Each player randomly chooses to switch (s) or not (ns) his initial choice. This information is then made public for $\mathcal{P}_{1}$ to have it.
	
	\item Based on the information from the previous step, $\mathcal{P}_{1}$ applies or not the corresponding door-switching operators $\hat{S}_{k}$.
	
	\item $\mathcal{P}_{1}$ measures the $n$ qudits corresponding to the labels in $\vec{p}$, and publicly announces which players won (w) and which ones lost (l).
	
	\item Knowing the game results, the players who had switched (s) and won (w), or not switched (ns) and lost (l), negate their bit $b_{k}$ from step 3.
\end{enumerate}

\begin{figure}[!t]
\centering
\includegraphics[width=1.0\linewidth]{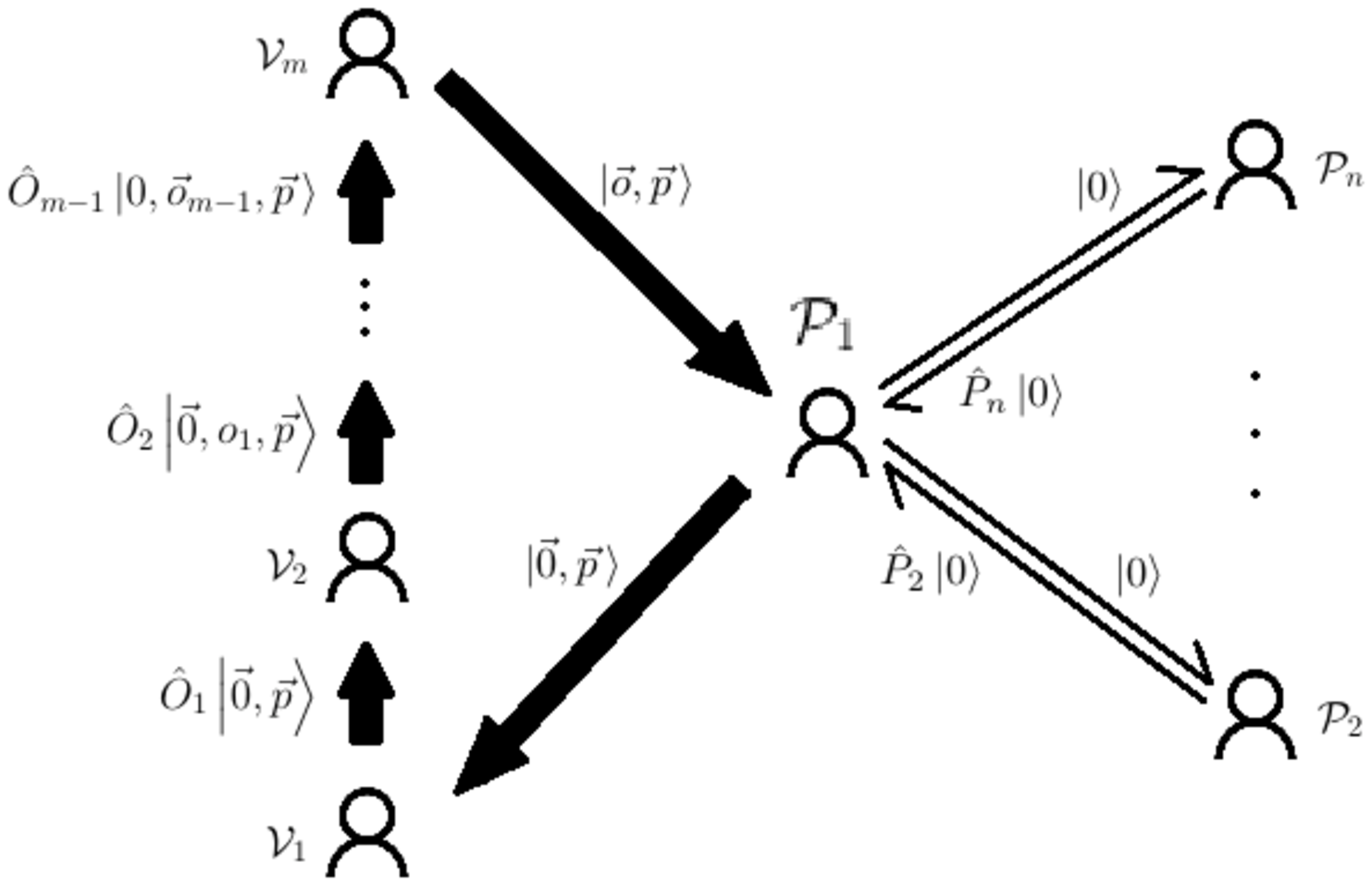}
\caption{Scheme describing a validated, multi-party, key-distribution, quantum protocol. The validation procedure is protected from eavesdropping via the entanglement of the remaining state. \label{Scheme}}
\end{figure}

Figure \ref{Scheme} shows a scheme of the proposed protocol. The protocol works due to the gap-filling function of the door-opening operators. At the beginning of the game, all the labels $p_{k} = 0$, then, after the host and all players have applied their strategy $p_{k} \in \left\lbrace 0,1 \right\rbrace $. The $ m = d-2$ door-opening operators fill all the gaps from $2$ to $d-1$, making possible for the door-switching operators to jump from $d-1$ to $0$. For each player $\mathcal{P}_{k}$ there are exactly eight possible results:
\begin{align}
	&\left( b_{1} = 0 , b_{k} = 0 , \text{s} , \text{l} \right), \notag \\
	&\left(  b_{1} = 0 , b_{k} = 0 , \text{ns} , \text{w} \right), \notag \\
	&\left( b_{1} = 0 , b_{k} = 1 , \text{s} , \text{w} \right), \notag \\
	&\left(  b_{1} = 0 , b_{k} = 1 , \text{ns} , \text{l} \right), \notag \\
	&\left(  b_{1} = 1 , b_{k} = 0 , \text{s} , \text{w} \right), \notag \\
	&\left( b_{1} = 1 , b_{k} = 0 , \text{ns} , \text{l} \right), \notag \\
	&\left(  b_{1} = 1 , b_{k} = 1 , \text{s} , \text{l} \right), \notag \\
	&\left(  b_{1} = 1 , b_{k} = 1 , \text{ns} , \text{w} \right).
\end{align}
Notice that the cases switched (s) and won (w), and not switched (ns) and lost (l) are precisely the ones in which the host and the player did not apply the same strategy (i.e. $b_{1} \neq b_{k}$), and hence the need of step 11. At the end of the protocol, if all the validation parties applied their corresponding door-opening operator, the host and all players will share the same value on their bit $b_{k}$, and the game can be repeated depending on the desired length of the key. Furthermore, since the validation part of the game state remains after the measurements performed in step 10, and this remaining state is entangled, even though not maximally, it can be used in a Bell-type test to ensure the validation process was done without external intervention.

It is worth mentioning that there exist specific cases in which the protocol will not work, namely, when $p_{1} = p_{2} = \cdots = p_{n}$. This is satisfied when all $\mathcal{P}_{1},\dots,\mathcal{P}_{n}$ applied the same strategy, that is, either $b_{1} =  \cdots = b_{n} = 0$ or $b_{1} = \cdots = b_{n} = 1$. However, the probability of the host and all players applying the same strategy is $1/2^{n-1}$, which decreases exponentially with the number of players $n$ and, thus, can be consider as an unlikely error for large networks.

It should be mentioned that, in order for all the players $\mathcal{P}_{1},\dots,\mathcal{P}_{n}$ to share the same key at the end of the protocol, it is necessary that all validation parties $\mathcal{V}_{1},\dots , \mathcal{V}_{m}$ apply their corresponding door-opening operator, that is, the validation parties function is to approve the key distribution. The criteria for approval would of course depend on the specific scenario where the protocol is used, it may be some background information regarding the players or simply the will of a person.

\subsection{Motivated application} \label{motivated}

Motivated by some of the key features of the quantum version of the generalized multi-player Monty Hall game, in this subsection we propose another validated multi-party quantum key-distribution protocol. Unlike the one proposed in the previous subsection, where the validation procedure is the one protected by the remaining state's entanglement, in this protocol validation itself plays a secondary role, and the distribution of the key is the one protected by entanglement.

The protocol is developed under the conditions $d = m + 2 = n + 1$ and goes as follows:

\begin{enumerate}
	\setcounter{enumi}{0}
	\item $\mathcal{P}_{1}$ generates the state
	\begin{equation}
		\Ket{\psi} = \Ket{\vec{0}} \otimes \sum_{i=0}^{d-1} \, \ket{\underbrace{i \cdots i}_{n}},
	\end{equation}
	i.e. the state where all the $m$ labels in $\vec{o}$ are equal to zero, and the players' qudits $\vec{p}$ are prepared in the GHZ state of dimension $d$ between $n$ parties.
	
	\item $\mathcal{P}_{1}$ sends the qudits $p_{2},\dots,p_{n}$ to the respective $n-1$ players $\mathcal{P}_{2},\dots,\mathcal{P}_{n}$, and keeps the qudits $p_{1}$ and all the ones in $\vec{o}$.
	
	\item Each participant $\mathcal{P}_{k}$ ($k=1,\dots,n$) randomly applies, to its own qudit, one of the following two strategies: $\hat{P}_{k} = \text{SUM}_{d}(0)$, $\hat{P}_{k} = \text{SUM}_{d}(1)$. And privately stores in a classical bit $b_{k}$ its selection: $b_{k} = 0$ for $\text{SUM}_{d}(0)$ and $b_{k} = 1$ for $\text{SUM}_{d}(1)$.
	
	\item All the players send back the qudits $p_{2},\dots,p_{n}$ to $\mathcal{P}_{1}$.
	
	\item $\mathcal{P}_{1}$ sends qudits $p_{j}$ and $o_{j-1}$ to the corresponding validation party $\mathcal{V}_{j-1}$ for all $j \in \left\lbrace 2,\dots,n \right\rbrace $.
\end{enumerate}

One of the key features of the door-opening operators defined in equation \eqref{open}, is the gap-filling property discussed in Subsec. \ref{ent}. Here, we define a simpler variation of the door-opening operators that preserves the gap-filling property without creating the superposition of all possible combinations of opened doors, namely
\begin{equation}
	\label{open2} \hat{\Omega}_{j} = \sum^{d-1}_{i=0} \Ket{i \oplus j,i} \Bra{0,i}.
\end{equation}
where $j\in \left\lbrace 2,\dots,n \right\rbrace $. Operators $\hat{\Omega}_{j}$ act on the space corresponding to the labels $o_{j-1}$ and $p_{j}$, and are special unitary in their domain (states with $o_{j-1} = 0$).

\begin{enumerate}
	\setcounter{enumi}{5}
	\item For all $j \in \left\lbrace 2,\dots,n \right\rbrace$, validation party $\mathcal{V}_{j-1}$, based on some authentication or relevant information related to the transaction, applies (or not) its own variation of the door-opening operator $\hat{\Omega}_{j-1}$, and sends back qudits $o_{j-1}$ and $p_{j}$ to $\mathcal{P}_{1}$.
		
	\item Each player randomly chooses to switch (s) or not (ns) his initial choice. This information is then made public for $\mathcal{P}_{1}$ to have it.
	
	\item Based on the information from the previous step, $\mathcal{P}_{1}$ applies or not the corresponding door-switching operators $\hat{S}_{k}$.
\end{enumerate}

In order to know which players won (w) and which ones lost (l), without measuring their qudits (which would make the initial entanglement of the GHZ state disappear), we use the same strategy as in \cite{QGA4}. That is, we define victory-encoding operators $\hat{V}_{j}$, whose function is to encode in qudit $o_{j-1}$ the result of player $\mathcal{P}_{j}$:
\begin{equation}
	\label{victory} \hat{V}_{j} = \sum^{}_{i,k} \left( \delta_{i,k \ominus 1} + \delta_{i,k} + \delta_{i,k \oplus 1} \right) \cdot \Ket{\left| k-i \right|, i, k} \Bra{i \oplus j, i, k},
\end{equation}
where the sum runs form $0$ to $d-1$, $\delta_{i,k}$ stands for Kronecker's delta and $j\in \left\lbrace 2,\dots,n \right\rbrace $. Operators $\hat{V}_{j}$ act on the space corresponding to the labels $o_{j-1}$, $p_{j}$ and $p_{1}$, and are special unitary in their domain (states with $p_{j} = p_{1} \ominus 1$, $p_{j} = p_{1}$ or $p_{j} = p_{1} \oplus 1$).

Notice that, after the application of the victory-encoding operator $\hat{V}_{j}$, the case where player $\mathcal{P}_{j}$ won ($p_{j} = p_{1}$) results in $o_{j-1} = 0$, while the case in which $\mathcal{P}_{j}$ lost $p_{j} \neq p_{1}$ results in $o_{j-1} = 1$.

\begin{enumerate}
	\setcounter{enumi}{8}
	\item $\mathcal{P}_{1}$ applies the $n-1$ victory encoding operators $\hat{V}_{j}$ to the corresponding qudits.
		
	\item $\mathcal{P}_{1}$ measures the $m$ qudits corresponding to the labels in $\vec{o}$, and publicly announces which players won (w) and which ones lost (l).
	
	\item Knowing the game results, the players who had switched (s) and won (w), or not switched (ns) and lost (l), negate their bit $b_{k}$ from step 3.
\end{enumerate}

\begin{figure}[!t]
	\centering
	\includegraphics[width=1.0\linewidth]{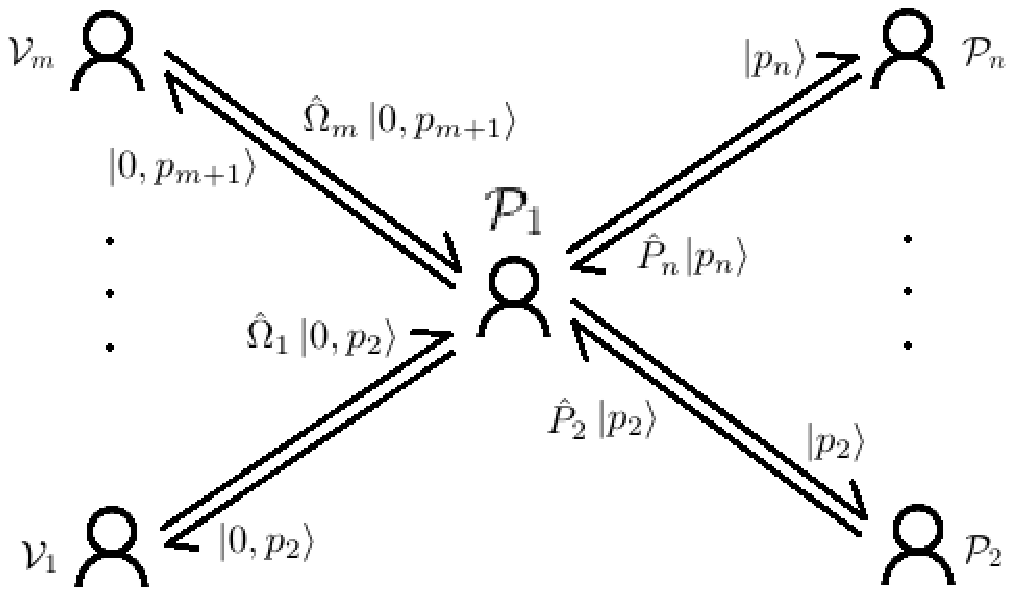}
	\caption{Scheme describing a validated, multi-party, key-distribution, quantum protocol. The key-distribution procedure is protected from eavesdropping via the entanglement of the remaining state. \label{Scheme2}}
\end{figure}

Figure \ref{Scheme2} shows a scheme of the protocol proposed in this subsection. The protocol works due to the gap-filling property of the operators $\hat{\Omega}_{j}$, while its security relies on the encoding function of the victory-encoding operators $\hat{V}_{j}$. As in the protocol of the previous subsection, for each player $\mathcal{P}_{k}$ there are exactly eight possible results:
\begin{align}
	&\left( b_{1} = 0 , b_{k} = 0 , \text{s} , \text{l} \right), \notag \\
	&\left(  b_{1} = 0 , b_{k} = 0 , \text{ns} , \text{w} \right), \notag \\
	&\left( b_{1} = 0 , b_{k} = 1 , \text{s} , \text{w} \right), \notag \\
	&\left(  b_{1} = 0 , b_{k} = 1 , \text{ns} , \text{l} \right), \notag \\
	&\left(  b_{1} = 1 , b_{k} = 0 , \text{s} , \text{w} \right), \notag \\
	&\left( b_{1} = 1 , b_{k} = 0 , \text{ns} , \text{l} \right), \notag \\
	&\left(  b_{1} = 1 , b_{k} = 1 , \text{s} , \text{l} \right), \notag \\
	&\left(  b_{1} = 1 , b_{k} = 1 , \text{ns} , \text{w} \right).
\end{align}
Notice again that the cases switched (s) and won (w), and not switched (ns) and lost (l), are precisely the ones in which the host $\mathcal{P}_{1}$ and the player $\mathcal{P}_{k}$ did not apply the same strategy (i.e. $b_{1} \neq b_{k}$), and hence the need of step 11. At the end of the protocol, if all the validation parties applied their corresponding variation of the door-opening operator, the host and all players will share the same value on their bit $b_{k}$, and the game can be repeated depending on the desired length of the key. Furthermore, after the measurements performed in step 10, the remaining state, corresponding to the players' labels in $\vec{p}$, is maximally entangled in a state equivalent to the GHZ state, and thus it can be used in a Bell-type test to ensure the key distribution was done without external intervention.

It should be mentioned that, as in the protocol of the previous subsection, in order for all the players $\mathcal{P}_{1},\dots,\mathcal{P}_{n}$ to share the same key at the end of the protocol, it is necessary that all validation parties $\mathcal{V}_{1},\dots , \mathcal{V}_{m}$ apply their corresponding variation of the door-opening operator. However, unlike the previous protocol, here the relevant feature is the key distribution, and thus the validation procedure (steps 5 and 6) can be completely omitted, simply by allowing the host $\mathcal{P}_{1}$ to apply the operators $\hat{\Omega}_{j}$. It is also worth mentioning that the protocol propose in this subsection will not work when $p_{1} = p_{2} = \cdots = p_{n}$. But, since the probability of this to happen is $1/2^{n-1}$, which decreases exponentially with the number of players $n$, it may be considered as an unlikely error for large networks.

\section{Discussion and Conclusions} \label{Conclusions}

In this work we developed a quantum version of a generalization of the Monty Hall game. In the case where a non-entangled initial state is used, we recover the classical expected payoff by allowing the host to hide the prize in a homogeneous superposition of doors, and the player to choose only one of them. We also showed that, by using a superposition of the switching and not-switching decision, it is possible to exceed the classical probabilities for the player to win. Furthermore, a quantum weak Nash equilibrium is found in a set independent of the host's strategy.

When an entangled GHZ state is used as the initial state of the game, the classical result can be again recovered via a destructive interference caused by the host and the player both using the quantum Fourier transform as their strategy. We also showed that a displacement of the correlation present in the GHZ state, depending on the parameters values, can lead to a very marked difference between the switching and not-switching cases.

In both the separable and entangled initial-state cases, the importance of the number of doors to be opened by the host plays a significant role in the player's expected payoff. In the separable case, it is decisive in the maximum value the payoff can have, while in the entangled case, the fewer doors are opened, the greater the cases in which the player cannot win regardless of his strategy. The heavy dependence of the player's payoff with this parameter, was the reason we decided not to treat it as a host's possible strategy, nonetheless this is a case we would like to address in a future work.

In the last section we extend our quantum scheme of the game to include multiple independent players, and use this extension to sketch two validated, multi-party, key-distribution, quantum protocols. The first protocol is proposed as a direct application of the multi-player quantum Monty Hall game, in the sense that it uses the same operators and mechanics. On the other hand, the second proposed protocol is less faithful to the game, as it only uses some of its key features to accomplish its purpose. Both protocols are protected by the entanglement of a remaining state, which can be used in a Bell-type test to ensure there was no external intervention in some specific steps: the validation procedure in the case of the first protocol, and the key distribution in the case of the second protocol.

We conclude that the mechanics of quantum games, or the quantum versions of some classical games, apart from being interesting from a basic-science perspective, can provide some useful insight in the search for a solution to multiple kinds of problems in applied quantum mechanics, particularly in the area of quantum information.

%\FloatBarrier

\acknowledgments
This work was supported by SEP-CONACYT under project no. 288856 and partially by 20200981-SIP-IPN, Mexico.

%\vspace{2cm}

\end{document}